\documentclass[aps,prl,groupedaddress,twocolumn]{revtex4-2}

\newcommand{\W}{{\cal W}}
\newcommand{\Md}{M_d}

\usepackage{amssymb}
\usepackage{amsmath}
\begin{document}

\newcommand{\HRule}{\rule{\linewidth}{0.5mm}}
\newcommand{\de}{\text{d}}
\newcommand{\pd}{\ensuremath{\partial}}
\newcommand{\iu}{\ensuremath{i}}
\newcommand{\eu}{\ensuremath{\text{e}}}
\newcommand{\AdS}{\ensuremath{AdS_d}}
\newcommand{\Cd}{\ensuremath{ C_d}}
\newcommand{\AdSC}{\ensuremath{AdS_d^{(c)}}}
\newcommand{\scalar}[2]{\ensuremath{#1\cdot #2}}
\newcommand{\lT}{\ensuremath{\mathcal Z_+}}
\newcommand{\rT}{\ensuremath{\mathcal Z_-}}
\newcommand{\lrT}{\ensuremath{\mathcal Z_\pm}}
\newcommand{\rlT}{\ensuremath{\mathcal Z_\mp}}
\newcommand{\xxi}{\zeta}

\def\RR{\mathbb R}
\def\C{\mathbb C}
\def\Bbb{\bf}
\newcommand{\f}{\underline{f}}
\newcommand{\xx}{{\mathbf{x}}}\newcommand{\yy}{{\mathbf{y}}}
\newcommand{\zz}{{\mathbf{z}}}
\newcommand{\cc}{{\mathbf{c}}}
\newcommand{\g}{\underline{g}}
\newcommand{\h}{\underline{h}}
\newcommand{\x}{\mathrm{x}}
\newcommand{\y}{\mathrm{y}}
\newcommand{\z}{{\rm{z}}}
\renewcommand{\k}{{\mathrm{k}}}
\newcommand{\p}{{\mathrm{p}}}
\newcommand{\zu}{z}

\def\K{\varkappa}

\def\Cb{{\Bbb C}}
\def\Rb{{\Bbb R}}
\def\W{\makebox{\Eul{W}}}

\font\Eul = eufm7 at 12pt \font\eul = eufm7 at 9pt
\renewcommand{\cosh}{\mbox{ch\,}}\renewcommand{\sinh}{\mbox{sh\,}}
\renewcommand{\tanh}{\mbox{th\,}}\renewcommand{\tan}{\mbox{tg\,}}
\def\dil{{D}}

\def\bC{{\bf C}}
\def\bR{{\bf R}}
\def\bN{{\bf N}}
\def\bZ{{\bf Z}}
\def\Im{{\rm Im\,}}
\def\Re{{\rm Re\,}}
\def\Rp{{\bf R}_+}
\def\Rm{{\bf R}_-}
\def\bCp{{\bf C}_+}
\def\bCm{{\bf C}_-}
\def\tg{{\rm tg\,}}
\def\th{{\rm th\,}}
\def\ch{{\rm ch\,}}
\def\sh{{\rm sh\,}}
\def\AA{{\cal A}}
\def\BB{{\cal B}}
\def\CC{{\cal C}}
\def\DD{{\cal D}}
\def\EE{{\cal E}}
\def\FF{{\cal F}}
\def\GG{{\cal G}}
\def\HH{{\cal H}}
\def\II{{\cal I}}
\def\JJ{{\cal J}}
\def\KK{{\cal K}}
\def\LL{{\cal L}}
\def\MM{{\cal M}}
\def\NN{{\cal N}}
\def\OO{{\cal O}}
\def\PP{{\cal P}}
\def\QQ{{\cal Q}}
\def\RR{{\cal R}}
\def\SS{{\cal S}}
\def\TT{{\cal T}}
\def\UU{{\cal U}}
\def\VV{{\cal V}}
\def\WW{{\cal W}}
\def\XX{{\cal X}}
\def\YY{{\cal Y}}
\def\ZZ{{\cal Z}}
\def\wh{\widehat}
\def\wt{\widetilde}
\def\ovl{\overline}
\def\unl{\underline}
\def \vhi{\varphi}
\def \veps{\varepsilon}
\def\e{{\rm e}}
\def\HB{\hfill\break}
\def\interior#1{\setbox1=\hbox{$#1$}\rlap{$#1$}\kern0.4\wd1\raise1.1\ht1%
\hbox{$\scriptstyle \circ$}}
\def\bydef{\mathrel{\buildrel \hbox{\scriptsize \rm def} \over =}}
\def\boxit#1#2{\setbox1=\hbox{\kern#1{#2}\kern#1}%
\dimen1=\ht1 \advance \dimen1 by #1 \dimen2=\dp1 \advance \dimen2 by #1
\setbox1=\hbox{\vrule height\dimen1 depth\dimen2\box1\vrule}%
\setbox1=\vbox{\hrule\box1\hrule}%
\advance \dimen1 by .4pt \ht1=\dimen1 \advance \dimen2 by .4pt \dp1=\dimen2
\box1\relax}
\def\endprf{\raise .5ex\hbox{\boxit{2pt}{\ }}}

\def\Rep{\mbox{$\phi$}}
\def\amb{{\bf R}_{2}^{d+1}}
\def\ambc{{\bf C}_{2}^{d+1}}
\def\wTp{{\wt \TT_{+}}}
\def\wTm{{\wt \TT_{-}}}
\def\wTpm{{\wt \TT_{1\pm}}}

\def\ifundefined#1{\expandafter\ifx\csname#1\endcsname\relax}

\title{\bf Anti-de Sitter, plane waves and quantum field theory}
\author{Ugo Moschella}
\address{Dipartimento di Scienza e Alta Tecnologia, Via Valleggio 11, 22100 Como,
and INFN sez. di Milano, Italy}
\address{Institut des Hautes Etudes Scientifiques, 35 Route de Chartres, 91440 Bures-sur-Yvette, France}
\date{\today}

\begin{abstract}
We present a new plane-wave expansion of the Wightman functions of anti de Sitter scalar fields and showcase its conceptual and technical importance in AdS quantum field theory. We deduce from it a new integral representation of the Feynman propagator which helps in clarifying  the relation between AdS Euclidean and Lorentzian Feynman diagrams in concrete examples. The plane-wave  expansion makes it possible also to demonstrate numerous new nontrivial formulas for Bessel and Legendre functions, and we provide two examples.
\end{abstract}
\maketitle
It is hard to overstate the importance of having manifestly covariant momentum space representations of relativistic quantum fields based on plane waves.  Such representations provide  the physically intuitive ingredients of Feynman diagrams and a great deal of the predictive power of Quantum Field Theory (QFT) depends on them. Plane wave expansions do exist  also  for de Sitter (dS) quantum fields \cite{bgm,bm}  while they still do not in the  anti-de Sitter (AdS) case, which may come as a surprise after 50 years of research in AdS.  The challenge cannot be tackled within  the usual approach to AdS QFT that only focuses on the boundary conditions at spacelike infinity \cite{avis,breit} but largely ignores the global analyticity properties of the AdS manifold  and of the correlation functions of the fields \cite{bem}.  Disregarding those properties, the topological difficulties associated with the existence of closed timelike curves and the necessity to move to the covering of the manifold 
cannot be overcome. 
Here we present a solution for AdS scalar fields in  spacetime dimension $d$ and open a new way to deal with AdS QFT.
 Tensorial and spinorial correlation functions may be derived by applying suitable differential operators to our formulae.

The dS and the AdS spacetimes are real Lorentzian submanifolds of the  complex sphere $S_d^{(c)}$  having quite  different properties: $dS_d$ is globally hyperbolic but does not admit global timelike Killing vector fields while the contrary happens for  $\AdS$.
These features make the corresponding QFT's very different too, but, as  we will show in this letter, the manifestly covariant plane-wave expansions of the Wightman functions are almost identical, the main difference being of a topological nature.

Let us start by recalling   the structure of the two-point function of a scalar field in a $d$-dimensional Minkowski spacetime $\Md$ with  metric  $\eta_{\mu\nu}=(+,-,\ldots,-)$: 
\begin{eqnarray}\label{tp}
&& {\cal W}^{M_d}_{m}(\x_1,\x_2) =
\int  \Psi^{(-)}_{\vec \p} (\x_1)  \Psi^{(+)}_{\vec \p} (\x_2)d\vec \p, \label{aa}\\ 
&& \Psi^{(\pm)}_{\vec \p} (x) = \frac {\exp(\pm i \p\cdot \x)}{\sqrt {2(2\pi)^{d-1} \omega}}, \  \p^0=\sqrt{|\vec{\p}|^2+m^2} \label{bb}.
\end{eqnarray}
The integral  (\ref{aa}) is the superposition of a complete set  of {\em plane waves} (labeled by all possible $(d-1)$-momenta $\vec{p}$)  evaluated at a first point $x_1$ multiplied by their complex conjugates evaluated at a second point $x_2$: it has a  distributional meaning but it invites us to move the first point into the backward tube  $T_-$ and the second point into the forward tube $T_+$ of the complex Minkowski space $\Md^{(c)}$:
\begin{eqnarray}
&& {T}_\pm= \{\z = \x+i \y \in \Md^{(c)}: \ \pm \y \in V_+\}, \\
&& {V}_+= \{ \y \in \Md: \   \y\cdot \y >0,\  \y^0 >0\}.
\end{eqnarray}
Since the  waves are exponentially decreasing in the respective tubes, this move vastly improves the convergence of the integral 
and shows that the {\em distribution} (\ref{aa}) is one of a very special kind:  it is  the boundary value on the reals of a {\em function} $W^{\Md}_{m}(\z_1,\z_2)$  
 holomorphic in the {\em domain} 
 $T_-\times T_+$, a mathematical property  equivalent to the  physical requirement that the  states of the theory have positive energy  in every inertial frame;  
such fundamental equivalence  holds also for the $n$-point functions of interacting fields \cite{pct} and is so significant that the Euclidean formulation of QFT would not exist without it \cite{1}. 

The (complex) $d$-dimensional de Sitter space-time
\begin{eqnarray}
dS_d^{(c)} = \{z \in M_{d+1}^{(c)}\ :\ z\cdot z= \eta_{\mu\nu} z^\mu z^\nu = -R^2\}
\label{s.2.1}\end{eqnarray}
also contains dS invariant   {\em domains of analyticity} \cite{bm} 
 describable simply as the intersections of $dS_d^{(c)}$ with  the  tubes $T_\pm$ of the ambient complex Minkowski space $M_{d+1}^{(c)}$:
\begin{eqnarray}
{\cal T}_\pm =   dS_d^{(c)} \cap T_\pm= \{x+iy \in dS_d^{(c)}\ :\ y\in \pm V_+\}.
\label{s.2.1}\end{eqnarray}
The second observation is that the asymptotic lightcone 
\begin{equation}
    C_+= \partial V_+= \{\xi \in M_{d+1}, \ \ \xi^2= 0, \ \ \xi^0>0\},
    \end{equation}
which describes the boundary at timelike infinity of the real de Sitter manifold, may be  interpreted as the space of momentum directions;  the fact that timelike geodesics and conserved quantities can be parametrized by the choice of  two such vectors confirms this interpretation:
\begin{eqnarray}
&& x^{\mu} (\tau)   = \frac{R (\xi^\mu e^{\frac{\tau}R}- \eta^\mu e^{-\frac{\tau}R})}{\sqrt{2 \xi\cdot \eta}}, \label{geo} \ K^{\mu\nu}= \frac{m({\xi\wedge \eta })^{\mu\nu}}{({\xi\cdot \eta})}
\end{eqnarray}
 where $\tau$ is the proper time and $m$ is the mass of the classical particle in geodesic motion \cite{cacc}.

What replaces the waves (\ref{bb}) in the de Sitter universe? Given  any complex number $\lambda$ (the mass parameter),  
dS plane waves \cite{bm,bgm} are constructed as follows 
\begin{equation}
 {\cal T}_\pm \times C_+ \ni z,\xi  \to \psi^{\pm}_\lambda(z,\xi) = 
\left({  \xi\cdot z}\right)^{\lambda}  =  e^{\lambda \log \left({ \xi\cdot  z}\right) }.   \label{dswaves}
\end{equation}
These functions are univalued and holomorphic in the tubular domains $\cal T_-$ and $\cal T_+$ and  solve  there the Klein-Gordon equation
$ (\Box_{dS}    +\mu_\lambda^2) \psi^\pm _\lambda (z) =0 \label{KGnu}$
with complex squared mass 
$
\mu^2_\lambda = \lambda(1-d-\lambda). 
$
The  ``phase'' of the above waves  is constant on the planes $z\cdot \xi = const$.

For any pair $z_1,z_2$ of points in the domain ${\cal T}_-\times  {\cal T}_+$ the canonically normalized (so-called) Bunch-Davis dS two-point function   (here generalized to complex squared masses and to spacetime dimension $d$) is a holomorphic superposition of plane waves  similar to (\ref{aa}):
\begin{eqnarray}
 W^{dS_d}_{\lambda}(z_1,z_2)    = \frac{ e^{ i \pi\left(\lambda+\frac{ d-1}{2} \right) } \Gamma (-\lambda ) \Gamma
   (\lambda+d-1)}{2^{d+1} \pi ^{d}} \cr  \times \int_{\gamma} \psi^{-}_\lambda(z_1,\xi)\, \psi^{+}_{1-d-\lambda}(z_2,\xi)\,d\mu_\gamma(\xi)  \label{uup} 
\end{eqnarray}
where $ d\mu_\gamma(\xi)$
is the restriction of the volume form of the cone $C_+$ 
to a complete ($d-1$)-dimensional cycle  $\gamma$. 

{\em Crucial property} :  the homogeneity degree $(1-d)$ of the integrand 
implies that the rhs is the integral of a  closed  differential form and therefore it is manifestly de Sitter invariant; it thus depends only on the invariant scalar product $z_1\cdot z_2$; a simple calculation  \cite{bgm,bm} shows that it is proportional to a  Legendre function of the first kind: 
\begin{eqnarray}
W^{dS_d}_{\lambda} (z_1,z_2)  = {1\over 2 (2\pi)^{d/2}} \Gamma(-\lambda)\Gamma(\lambda+d-1)\times \cr \times { \,    ((z_1\cdot z_2)^2-1)^{{2-d \over 4}}}  P_{ \lambda +{d-2\over 2} }^{-{d-2 \over 2}}(z_1\cdot z_2). \label{wig}
\end{eqnarray}
Formula (\ref{wig}) explicitly provides  the  analytic continuation of the two-point function to the cut-plane
\begin{equation}
    \Delta_{dS} =  \{ z_1,z_2 \in dS_d^{(c)}:\    z_1\cdot z_2 \not= \rho ,\ \   \rho \leq -1\}\end{equation} 
    which contains all pairs of complex events of $dS_d^{(c)}$ minus the causal cut; it   contains in particular all the non-coincident pairs of dS Euclidean points. Eq. (\ref{wig}) also shows that the involution 
$
\lambda\to (1-d-\lambda) \label{inv}
$ which leaves $\mu^2_\lambda$ invariant 
is a symmetry of the  two-point function.
\vskip 10pt

The complex  $d-$dimensional AdS  manifold can   also be defined by its embedding, now in the complex space  $\ambc$ which is a copy of $ {\bf C}^{d+1}$ endowed with the scalar product
\begin{equation}
    z_1\cdot z_2 = 
z_1^0z_2^0-z_1^1z_2^1-\ldots- z_1^{d-1}z_2^{d-1}+ z_1^{d}z_2^{d} \label{scalar}\end{equation} (in the following  the dot refers to this product):
\begin{equation}
\AdSC=\{z \in \ambc,\ z^2=z\cdot z = R^2\}.
\end{equation}
The real null cone of the ambient space 
\begin{equation}
\CC_d=\{\xi \in \amb,\ \xi^2=\xi\cdot \xi = 0, \ \xi\not=0\}
\end{equation}
describes the boundary at spacelike infinity of the real manifold $\AdS$; real null vectors may here be interpreted as {\em spacelike} momentum directions and used to parametrize spacelike AdS geodesics as in Eq. (\ref{geo}). Timelike geodesics and  the  associated conserved quantities are instead parametrized by null  vectors of the complex cone
\begin{eqnarray}
&& \CC_d^{(c)}=\{\xxi \in \ambc,\ \xxi\cdot \xxi =0, \ \xxi \not=0 \}
\end{eqnarray}
$$ x^\mu(\tau) =  \frac{R\,\xxi^\mu e^{\frac{i  \tau}R} + {R\, \xxi^\mu}^* e^{-\frac{i  \tau}R}}{\sqrt{2(\xxi \cdot \xxi^*)}},  \
K^{\mu\nu}  = \frac{{i\, m( \zeta\wedge   {\zeta}^* )^{\mu\nu}}}{ (\zeta \cdot \zeta^*)}.$$
This parametrization  shows that timelike geodesics which contain an event $x$ 
also contain the antipodal event $-x$ and that  the proper time elapsed between them is $\pi R$ on each geodesic.  
To remove the periodicity (but not the periodical focusing - see the  discussion given in \cite{gibbons})  of the geodesics and  to have well-defined plane waves it is necessary to move to the covering manifold.

In spacetime dimension  $d \ge2$ the covering of the complex manifold $\AdSC$ 
is $\AdSC$ itself. On the contrary, the real manifold $\AdS$ admits  a nontrivial 
{covering space}  
$\widetilde {\AdS}$; 
also the backward and forward {\em chiral} tuboids \cite{bem}
\begin{eqnarray}
\ZZ_{\pm} = \{z=x+iy \in \AdSC : 
\, \scalar{y}{y}  >0,\,  \cr  \epsilon(z)=  y^0 x^d - x^0 y^d \gtrless 0  \}  
\end{eqnarray} 
have nontrivial coverings $\widetilde \ZZ_{\pm}$.
Here again,  as in Minkowskian QFT,   any local and covariant two-point distribution with  positive energy  spectrum 
is the boundary value on $\widetilde \AdS\times \wt \AdS $ of a function  holomorphic in the  domain
$\wt{\cal Z}_{-}\times \wt{\cal Z}_+$.
These properties  completely  determine the two-point functions \cite{bem} and, as a consequence, select the possible  boundary behaviour of the modes. Analyticity  properties of the same kind characterize also the $n$-point functions of interacting fields and allow for the Euclidean formulation of AdS QFT \cite{bem}.

The last ingredients that we need 
are  the  two AdS-invariant  \textit{chiral cones}:    
\begin{eqnarray} \label{chiralcones+}
 \CC_{\pm} = \{\zeta=\xi+i\chi \in \CC_d^{(c)}:\
\scalar{\chi}{\chi}  >0,  \  \epsilon(\zeta) \gtrless 0 \}
\end{eqnarray}
and their punctured closures 
\begin{eqnarray} \label{chiralcones+}
\ovl\CC_{\pm} = \{\zeta\in \CC_d^{(c)}:\,
\scalar{\chi}{\chi}  \geq0,  \,  \epsilon(\zeta) \gtreqless 0,\, \zeta\not=0\}.
\end{eqnarray}
The scalar product
 $z,\zeta \mapsto \scalar{z}{\zeta}$  maps $\ZZ_{-} \times {\ovl\CC_{+}}$  and $\ZZ_{+} \times
{\ovl\CC_{-}}$ onto the punctured complex-plane  ${\bf C} ^*= {\bf C}\setminus \{0\}$: in the above domains the scalar product $\scalar{z}{\zeta}$ 
cannot vanish. Furthermore, the scalar product   can be lifted to a map of 
 $\widetilde\ZZ_{-} \times \ovl\CC_{+}$  and $\widetilde\ZZ_{+} \times \ovl\CC_{-}$ onto
$\widetilde{\bf C}^*$ which also never vanish. By abuse of language we denote also the lifted scalar product   by $ \scalar{z}{\zeta}$.

We are now in position to answer the question: what replaces the plane waves (\ref{bb})  in the AdS case? Given  any complex number $\lambda$, we introduce (unnormalized) AdS plane waves  as follows:  
\begin{equation}
\wt \ZZ_\pm \times \ovl \CC_\mp \ni \zeta,z \to \phi^{\pm}_\lambda(z,\zeta) = 
\left({ z\cdot \zeta}\right)^{\lambda}  =  e^{\lambda \log \left({ z\cdot \zeta}\right) }.   \label{waves}
\end{equation}
A previous attempt \cite{pene1} was based on a local $i \epsilon$-prescription. Here the waves are, in their global domains of definition, univalued, holomorphic  solutions of  the  KG equation 
$
(\Box_{AdS} + m_\lambda^2)\phi^{\pm}_\lambda(z,\xi)= 0  \label{cadskg}
$ with complex  squared mass
$
    m^2_\lambda =-\mu^2_\lambda =  \lambda(\lambda+d-1).$

Let us now consider a family of two-point functions formally identical to (\ref{uup}): 
\begin{equation}
 W^{AdS_d}_{\lambda }(z_1,z_2)   =c_d(\lambda)  \int_{\gamma} \phi^{-}_\lambda(z_1,\xi)\, \phi^{+}_{1-d-\lambda}(z_2,\xi)\,d\mu_\gamma(\xi).  \label{uup2}\end{equation}
When the  integral is extended to a complete cycle $\gamma$ of the {real} cone $\CC_d$ the  outcome  is not AdS invariant because the integrand gets multiplied by $\exp{(-4 \pi i \lambda)}$ after one revolution in the $(\xi^0,\xi^d)$-plane.  Exceptions are $\lambda=l$ integer, but the integral vanishes identically, and  $\lambda=l+\frac 12$ half-integer, which gives the correct AdS-invariant result. 
For generic $\lambda$, integrating on the covering of the cone does not  help and is physically unsound as it would  over-count the possible momentum directions. Indeed, a complete cycle $\gamma$ is already too much, and this partly explains the vanishing of the integral for integer $\lambda=l$. 

The solution for  general  $\lambda$ is given by Eq. (\ref{uup2}) with the following specifications: for $\Re \lambda >-1$ the cycle  $\gamma(z_1)$ should belong to a  relative homology class of $H_{d-1}(\CC_-,\{\zeta:\,\zeta\cdot z_1 =0\})$ (a similar construction based on the second point $z_2$ gives identical results);
the canonically normalized plane-wave expansion of the two-point Wightman function holomorphic in $\ZZ_-\times \ZZ_+$ is  given by 
\begin{eqnarray}
&& W^{\AdS}_{\lambda }(z_1,z_2)    = -\frac{  \pi ^{1-d} \Gamma
   (\lambda+d -1)}{ 2^{d+1}\cos \left(\frac{\pi  d}{2}\right) \Gamma (\lambda +1)} \times \cr
   &&  \label{pwads}  \times \int_{{\gamma(z_1)}}(z_1 \cdot \zeta)^\lambda\,(z_2\cdot\zeta)^{1-d-\lambda}\,d\mu_{\gamma}(\zeta) =\label{pwads} \\ &&
= {e^{-i\pi{d-2\over 2}} \over (2\pi)^{d\over 2}} ((z_1\cdot z_2)^2-1)^{\frac{2-d}4}
 Q_{{d-2\over 2}+\lambda}^{\frac{d-2} 2}(z_1\cdot z_2). \label{Q}
\end{eqnarray}
A symmetric presentation of (\ref{pwads}) is possible by using the shadow transformation of the waves  that follows from  (\ref{pwads}) by letting $z_2$ tend to the boundary (the cone $\CC_d$): 
\begin{eqnarray}
&& W^{\AdS}_{\lambda }(z_1,z_2)  =c'_d(\lambda)\int_{\gamma(z_1)} \int_{\gamma(z_2)}   (z_1\cdot \zeta_1)^{\lambda} \times \cr && \times \,   (\zeta_1\cdot \zeta_2)^{1-\lambda-d}
 ( z_2\cdot \zeta_2)^{\lambda} \, d\mu_{\gamma(z_1)} \, d\mu_{\gamma(z_2)}.
\end{eqnarray}
In odd spacetime dimension ($AdS_3$, $AdS_5$, \ldots)  Eq. (\ref{pwads}) has to be  understood as follows
\begin{eqnarray}
&& W^{AdS_{2n+1}}_\lambda(z_1,z_2)=
\frac{(-1)^{n+1} \Gamma(\lambda+2n
   )}{(2 \pi )^{d} \Gamma (\lambda +1)}\times \cr \times 
&& \int_{\gamma(z_1)} (z_1\cdot \zeta)^\lambda ( z_2\cdot \zeta)^{-\lambda-2n}\log(\zeta\cdot z_2) d\mu_\gamma(\zeta). \label{pwduodd} \
\end{eqnarray}
Eq. (\ref{Q}) provides the  analytic continuation of $W^{\AdS}_{\lambda} (z_1,z_2)$ both to the complex $\lambda$-plane and to the covering $\wt \Delta_{AdS}$ of the cut-plane
\begin{equation}
\Delta_{AdS} =  \{ z_1,z_2\in \AdSC; \,   z_1\cdot z_2\not= \rho ,\, -1\leq \rho \leq 1\}.
\end{equation}
$\Delta_{AdS}$ contains all pairs of  events of $\AdSC$ minus the causal cut; it   contains in particular all pairs of  non-coincident Euclidean points. Eq. (\ref{Q})  shows that 
$
\lambda\to (1-d-\lambda)$ 
is not a symmetry of  the two-point function.

Eq. (\ref{pwads}) is very significant for the study of Legendre functions of the second kind. To give one example, a nontrivial consequence is the, perhaps unknown,  multiplication theorem (calculated  at $d=2$; a more complicated formula of the same kind holds for general $d$): 
 \begin{eqnarray}
Q_\lambda (x\, \cosh u)= \sum_{n=0}^\infty\frac{\left(1-x^{-2} \right)^{ n} }{ x ^{1+\lambda }\Gamma (1+ n)  }    \,  
  \frac{Q_{n+\lambda }^n(\cosh u)}{  (2 \, \sinh u)^{n} } .\label{int01}
\end{eqnarray}
Let us consider now the Poincar\'e foliation of $\AdSC$:
\begin{equation}
z(\z,u)=\left\{\begin{tabular}{lcl}
 $z^{\mu} $ & =& $\frac{ 1}{u}\z^\mu  $\cr
 $z^{d-1} $& =& $\frac{1-u^2}{2u} +
\frac {1}{2 u} \z^2$  \cr
 $z^{d}$&= & $\frac{1+u^2}{2u} -
\frac {1}{2 u} \z^2$
\label{coordinates}
\end{tabular}\right., \, \z^2= \eta_{\mu\nu}\z^\mu\z^\nu.
\end{equation}
For real $u>0$ and real $\z^\mu= \x^\mu$ the above coordinates cover one half of $\AdS$ which we denote $AdS_M$. Its partial complexification ($u$ is kept real) contains the Euclidean manifold $EAdS_d$; 
 the leaves at fixed $u>0$ are copies of $M_{d-1}^{(c)}$; the Minkowskian tubes $T_\pm$ of the  leaves are contained into  $\ZZ_\pm$ (see \cite{bem} for more details).  The latter  property has also an important technical value and allows to compute the Fourier transforms of the waves  w.r.t. the  coordinates $\x^\mu$ and to derive, after some pain, a diagonal representation of the two-point function (see also \cite{bertola, bem}):
 \begin{eqnarray}
&& W^{\AdS}_\lambda(z_1(\z_1,u),z_2(\z_2,u'))=
\frac{\left(u u'\right)^{\frac{d-1}{2}}}{2 (2\pi) ^{d-2} } \times 
  \cr && \times \int \theta(\p^0) \theta(\p^2)  e^{-i \p(\z_1-\z_2)} J_{\nu}\left(u \sqrt{\p^2} \right) J_{\nu }\left(  u'\sqrt{\p^2}\right)d\p = \label{rt} \cr && = \frac{\left(u u'\right)^{\frac{d-1}{2}}}2 
  \int_0^\infty   {\rm W}^{M_{d-1}}_m(\z_1,\z_2) J_{\nu }(m u) J_{\nu }( m  u') dm^2 \label{rt2}
   \end{eqnarray}
   where $\nu(\lambda)=\frac{d-1}{2}+\lambda $ $ \p\,\z = \eta_{\mu\nu}\p^\mu\z^\nu$  and  $\p^2= \eta_{\mu\nu}\p^\mu\p^\nu$.
   The first equality is valid for $\z_1  \in T_-$ and $\z_2\in T_+$; the second equality   (\ref{rt})  has the form of a K\"all\'en-Lehmann  expansion and holds in the whole cut-plane 
   \begin{equation}
\Delta_M =  \{ \z_1,\z_2\in M_{d-1}^{(c)}; \ \   z_1\cdot z_2\not= \rho ,\ \   \rho \geq 0\}.\nonumber
\end{equation}
From Eq. (\ref{rt2})  it may be deduced a new integral representation of the Feynman propagator for pair of events $x_1,x_2$ {\em in  the real Poincar\'e chart} $AdS_M$:
   \begin{eqnarray}
&&G^{\AdS}_\lambda(x_1(\x_1,u),x_2(\x_2,u)) = \frac {\left(u u'\right)^{\frac{d-1}{2}}} 2
 \times  \cr &&  \times \, \int_0^\infty   {G}^{M_{d-1}}_m(\x_1,\x_2) J_{\nu }\left(m u  \right) J_{\nu}\left( m  u'\right) dm^2; \label{green}
\end{eqnarray}
here $G^{M_{d-1}}_m(\x_1,\x_2)$ is the Feynman propagator of massive Minkowskian scalar field in spacetime  dimension $d-1$: 
\begin{eqnarray}
&&(\Box_{AdS} +m^2_\lambda) G^{\AdS}_\lambda(x_1(\x_1,u),x_2(\x_2,u)) =\cr &&= 
 {\left(u u'\right)^{\frac{d-1}{2}}}
  \int_0^\infty  u^2 \delta(\z_1,\z_2) J_{\nu }\left(m u  \right) J_{\nu }\left( m  u'\right) m dm \cr && =u^d \delta(\z_1,\z_2) \delta(u,u') = \delta_{AdS_M}(x_1,x_2).
\end{eqnarray}
In the last equality  we used the inversion theorem for the Hankel transform of order $\nu$  \cite{batemantransform2} and $ \sqrt{g(x)} =u^{-d}$.

AdS Feynman Diagrams are usually computed in the Euclidean manifold, often by using the  "split"  representation of the Euclidean Feynman propagator \cite{ruhl,pene,pene0}: the latter  may be understood as the generalized  inverse Mehler-Fock transform  with respect to the mass parameter $\lambda$ of the {de Sitter} Wightman function (\ref{uup}) taken at purely imaginary events $z_1=iy_1\in \TT_-$  and $z_2=iy_2\in \TT_+$ \cite[Appendix A]{cem}.

The relation of AdS Euclidean diagrams with real space diagrams  \cite{pene1,pene,malda} remains however poorly understood, also because calculations based  on the split representation cannot be Wick-rotated.

Eq. (\ref{green})  may shed some new light on this issue. 
In particular it immediately implies that banana diagrams  computed in the Euclidean manifold $EAdS_d$ \cite{cem} can be Wick-rotated and coincide with the same diagrams calculated in $AdS_M$. 

What is less obvious is whether a more general diagram computed by integrating only in a Poincar\'e patch $AdS_M$  produces an AdS invariant result. This happens for the  elementary  one line diagram: 
\begin{eqnarray}
&&  \int_{AdS_M} G^{\AdS}_{\lambda}(x_1,x)  \sqrt{g(x)}\, dx =
\cr&& 
\int_0^\infty \int_0^\infty {\left( \frac{u'}{u}\right)^{\frac{d-1}{2}}}
    \frac{ J_{\nu}\left(m u  \right) J_{\nu }\left( m  u'\right)}{m u}    {du} dm . \nonumber \end{eqnarray}
Carrying the integration in the $u$ variable first we get
\begin{eqnarray}
&&   
\int_0^\infty \frac{m^{\frac{d-3}{2}} \Gamma
   \left(\frac{\lambda }{2}\right) {u'}^{\frac{d-1}{2}}  J_{\frac{d-1}{2}+\lambda }(m u')}{ 2^{\frac{d+1}{2}} \Gamma
   \left(\frac{1}{2} (d+\lambda +1)\right)}dm 
\label{diagram}=\cr &&=  \frac{1}{\lambda  (d+\lambda -1)}= \frac 1{m_\lambda^2}.\end{eqnarray}
The result does not depend on $u'$, is AdS invariant and   coincides with the Euclidean calculation. A less trivial example is the two-line diagram:
\begin{eqnarray}
F_{\lambda_1\lambda_2}(x_1,x_2) =\int_{AdS_M} G^{\AdS}_{\lambda_1}(x_1,x) G^{\AdS}_{\lambda_2}(x,x_2) \sqrt{g(x)} dx. \nonumber
\end{eqnarray}
Would we know the AdS invariance of $F_{\lambda_1\lambda_2}$ we could immediately write \cite{cem} 
\begin{eqnarray}
F_{\lambda\nu}(x_1,x_2) =-\frac{G^{\AdS}_{\lambda_1}(x_1,x_2) - G^{\AdS}_{\lambda_2}(x_1,x_2)}{m_{\lambda_1}^2-m_{\lambda_2}^2}. 
\end{eqnarray}
A direct calculation that confirms the above result may be performed by  Wick-rotating the identity 
\begin{eqnarray}
 \int_0^\infty   \frac{J_{\nu_1}\left(m u  \right) J_{\nu_1}\left(m u'  \right) - J_{\nu_2}\left(m u  \right) J_{\nu_2}\left( m  u'\right)}{({\nu_2^2-\nu_1^2})(\p^2+m^2)}
 m  dm = \cr     \int_0^\infty   \int_0^\infty   \int_0^\infty  \frac{ a b J_{\nu_1}\left(a u  \right) J_{\nu_1}\left( a  v\right)J_{\nu_2}\left(b v \right) J_{\nu_2}\left( b  u'\right)}{v\left(a^2+\p^2\right) \left(b^2+\p^2\right)}  {dv}  da db  \nonumber
\end{eqnarray}
that can be shown at first for $\p^2>0$ by using Eq. (\ref{Q}) and the Euclidean version of Eq. (\ref{green}). Taking the limit $x_1\to x_2$ we obtain as a consequence  the one-loop banana integral calculated in $AdS_M$, and this coincides with the Euclidean calculation done in  \cite{cem}. 

We may also compute Witten diagrams in real space: the external legs are the boundary values of the plane waves (\ref{waves}) and the propagator is once more the one given in Eq. (\ref{green}). "Ingoing" waves should be  boundary values of the waves  holomorphic in $\ZZ_-$  and "outgoing" waves boundary values of the waves holomorphic in $\ZZ_+$. 

We conjecture that also these diagrams integrated over $AdS_M$ give AdS invariant results (see also \cite{pene1,pene,malda}) but this is a question that remains at the moment open.

{\bf Acknowledgments.}  
I want to express my gratitude to Sergio Cacciatori for numerous suggestions. I would also like to thank the Institut des Hautes Etudes Scientifiques (Bures sur Yvette) and the Physics Department of Fudan University for their warm hospitality and support.

This letter is dedicated to the memory of Henri Epstein.


\begin{thebibliography}{999}


\bibitem{bgm}
J.~Bros, U.~Moschella and J.~P.~Gazeau,
Phys. Rev. Lett. \textbf{73}, 1746-1749 (1994)
\bibitem{bm}
J. Bros and U. Moschella,  
Rev. Math. Phys.   8, 327 ( 1996)
\bibitem{avis}
S.~J.~Avis, C.~J.~Isham and D.~Storey,
Phys. Rev. D \textbf{18}, 3565 (1978)
\bibitem{breit}
P.~Breitenlohner and D.~Z.~Freedman,
Annals Phys. \textbf{144}, 249 (1982)
\bibitem{bem}
J.~Bros, H.~Epstein and U.~Moschella,
Commun. Math. Phys. \textbf{231}, 481-528 (2002)
\bibitem{pct}
R.~F.~Streater and A.~S.~Wightman,
``PCT, spin and statistics, and all that,'' Princeton Univ. Press (2000)
\bibitem{1}
K.~Osterwalder and R.~Schrader,
Commun. Math. Phys. \textbf{31}, 83-112 (1973)
\bibitem{cacc}
S.~Cacciatori, V.~Gorini, A.~Kamenshchik and U.~Moschella,
Class. Quant. Grav. \textbf{25}, 075008 (2008)
\bibitem{gibbons}
G.~W.~Gibbons,
[arXiv:1110.1206 [hep-th]].
\bibitem{pene1}
M.~Gary, S.~B.~Giddings and J.~Penedones,
Phys. Rev. D \textbf{80}, 085005 (2009)
\bibitem{bertola}M.~Bertola, J.~Bros, V.~Gorini, U.~Moschella and R.~Schaeffer,
Nucl. Phys. B \textbf{581}, 575-603 (2000)
\bibitem{batemantransform2} Bateman, H. and Erdélyi, A. (1954) Tables of Integral Transforms, Vol. II. McGraw-Hill, New York.
\bibitem{ruhl}
T.~Leonhardt, W.~Ruhl and R.~Manvelyan,
J. Phys. A \textbf{37}, 7051 (2004)
\bibitem{pene} J. Penedones, 
JHEP 03
(2011), 025
\bibitem{pene0}
M.~S.~Costa, V.~Gon{\c{c}}alves and J.~Penedones,
JHEP \textbf{09}, 064 (2014)
\bibitem{cem}
S.~L.~Cacciatori, H.~Epstein and U.~Moschella,
JHEP \textbf{08}, 109 (2024)
\bibitem{malda} J.~Maldacena, D.~Simmons-Duffin and A.~Zhiboedov,
JHEP \textbf{01}, 013 (2017)
\end{thebibliography}
\end{document}